\documentclass[traditabstract]{aa}

\usepackage{color}

\usepackage{graphicx}
\usepackage{txfonts}
\usepackage{pstricks}
\include{epsf}
\usepackage{natbib}
\usepackage[hyphens]{url}
\usepackage{hyperref}  

\bibpunct{(}{)}{;}{a}{}{,}

\def\0{\phantom0}

\begin{document}

\title{A fast empirical method for galaxy shape\\ measurements in weak lensing surveys}

\titlerunning{MegaLUT}

\author{M.~Tewes\inst{\ref{inst1}} \and N.~Cantale\inst{\ref{inst1}} \and F. Courbin\inst{\ref{inst1}} \and T. Kitching\inst{\ref{inst2}} \and G. Meylan\inst{\ref{inst1}}}

\institute{Laboratoire d'astrophysique, Ecole Polytechnique F\'ed\'erale de Lausanne (EPFL), Observatoire de Sauverny, CH-1290 Versoix, Switzerland, \email{malte.tewes@epfl.ch} \label{inst1}
\and
SUPA, Institute for Astronomy, University of Edinburgh, Royal Observatory, Blackford Hill, Edinburgh, EH9 3HJ, U.K.\label{inst2}
}

\date{Received 26 March 2012 / Accepted 20 June 2012}
 
\abstract{We describe a simple and fast method to correct ellipticity measurements of galaxies from the distortion by the instrumental and atmospheric point spread function (PSF), in view of weak lensing shear measurements.
The method performs a classification of galaxies and associated PSFs according to measured shape parameters, and corrects the measured galaxy ellipticites by querying a large lookup table (LUT), built by supervised learning. We have applied this new method to the GREAT10 image analysis challenge, and present in this paper a refined solution that obtains the competitive quality factor of $Q = 104$, without any shear power spectrum denoising or training. Of particular interest is the efficiency of the method, with a processing time below 3 ms per galaxy on an ordinary CPU.}

\maketitle

\section{Introduction}
\label{intro}

Gravitational lensing offers a means to map the distribution of matter over a broad range of spatial scales. In the strong regime, gravitational lensing gives rise to multiple images of distant sources. This allows both to study lensed sources in great details  and to map the matter in the central parts of the lensing objects, either individual galaxies \citep[e.g.,][]{Bolton2008, Faure2008, Courbin2012} or galaxy clusters \citep[e.g.,][]{Coe2010, Shan2012}. In the weak regime, only one image of the source galaxy is seen and its ap\-parent distortion can only be measured statistically, by averaging the signal over many galaxies. This occurs either when the mass density in the lensing objects is too low, below the ``critical density'', or when the sources are separated from the lenses in projection on the plane of the sky. Strong and weak lensing are sometimes combined e.g., to probe simultaneously the core and the large scale halo of galaxy clusters \citep[e.g.,][]{Limousin2007}. 

On very large spatial scales, weak gravitational lensing is not caused anymore by mass along a specific line of sight, but rather by the combined gravitational fields of the large scale structures of the Universe. The signature of the lensing distortions, called {\it cosmic shear}, is best seen in this regime through its power spectrum or through its two-point correlation function across the whole sky. Since the first measurement of the effect  \citep{Maoli2001, Bacon2000, Kaiser2000, vanWaerbeke2000, Wittman2000}, it was quickly realized that cosmic shear is a sensitive tool to measure indirectly some of the most important cosmological parameters, including the dark energy equation of state parameter and its evolution with redshift \citep{Hu1999}. Several ground-based wide field surveys to measure cosmic shear with unprecedented accuracy are under way or under study (e.g., Pan-STARRS\footnote{\url{http://pan-starrs.ifa.hawaii.edu}}, DES\footnote{\url{http://www.darkenergysurvey.org}}, Subaru HSC\footnote{\url{http://www.naoj.org/Projects/HSC/}}, LSST\footnote{\url{http://www.lsst.org}}). Euclid\footnote{\url{http://www.euclid-ec.org}}, a space mission currently being implemented by ESA, will image at least 15,000 square degrees of space with one of the main science objectives being the measurement of cosmic shear \citep[see, e.g.,][]{Laureijs2011}. 

All the applications of gravitational shear, whether they be about galaxy and cluster halos or about dark energy, require the measurement of shapes of numerous and faint distant galaxies with optimal precision and without any significant systematic bias. Euclid will observe about 1.5 billion galaxies to achieve its scientific goal. However, any telescope produces images that are limited either by diffraction, by the Earth's atmosphere, or by both effects. The algorithms that will be used to measure galaxy shapes must correct for this smearing, characterized by the point spread function (PSF) of the entire image acquisition process. A considerable amount of work has been devoted so far to tackle this problem. Among the most popular approaches is the ``KSB'' family of methods \citep{KSB1995}, based on the measurement of the second order moments of the light distribution of galaxies. In these methods, the correction for the smearing by the PSF is done analytically. Many different implementations and improvements of KSB are currently in use. Other algorithms consider a fit of an analytical model to the galaxies \citep[e.g.,][]{Miller2007, Kitching2008} or decompose them on an orthogonal basis of vectors called ``shapelets'' \citep{Kuijken2006, Refregier2003a, Refregier2003b}. 

Even if the PSF is properly accounted for, galaxy shape measurements, as well as the resulting shear measurements, are possibly biased by the presence of noise in the images \citep[see e.g. ][]{Refregier2012, MelchiorViola2012}. It is likely that, given the complexity of galaxy shapes, this ``noise bias'' will have to be addressed by an empirical calibration using synthetic data \citep{Kacprzak2012}. Such calibrations can be performed at the topmost level of the shear measurement (i.e., the recovered shear itself), or at the lower levels of the shape measurements (i.e., correcting every galaxy measurement). Recently, \citet{gruen2010} introduced a promising method based on training a neural network to correct for the bias at the level of each individual galaxy. We propose in the present work an algorithm in line with the work of \citet{gruen2010}, but we apply a simple machine learning approach directly to the PSF removal problem instead of correcting for the residual bias of an existing PSF removal method.
This potentially yields unbiased galaxy shapes, which is the primary goal of this work. However, as emphasized by \citet{MelchiorViola2012}, also the \emph{variance} of shape estimates leads to a bias on the shear measurement. We show in this paper that our method is competitive even without any specific calibration of this noise bias at the level of the shear power spectrum.

The article is structured as follows : the principles of our method are described in Section \ref{principle}. Section~\ref{G10} describes an application to the simulated data of the GRavitational lEnsing Accuracy Testing 2010 (GREAT10) challenge \citep{GREAT10Handbook2010}, illustrating how our method can be combined with existing shape measurement techniques. The results achieved on GREAT10 are presented in Section~\ref{G10results}, while limitations and possible extensions to our method are discussed in Section~\ref{discussion}. Lastly, Section~\ref{conclusion} summarizes our conclusions.

\section{Description of MegaLUT}
\label{principle}

This paper proposes a conceptually simple, empirical, and fast method to correct ellipticity measurements of galaxies for the distortion by the instrumental and atmospheric PSF. The central idea of the method is to perform a classification of galaxy-PSF pairs with respect to their measured shape parameters, i.e., the parameters characterizing both the galaxy and its PSF.  For each of these classes, an ellipticity correction is estimated to remove the effect of PSF smearing. These corrections are obtained by supervised learning and written into a large but tractable lookup table (LUT), hence the name \emph{MegaLUT}. 
With this approach, the problem of correcting galaxy shapes for the convolution by the PSF is reduced to a simple array indexing operation.

The goal of MegaLUT is to describe at best the ellipticities of individual galaxies prior to the convolution by the PSF. We do not consider here the additional problem of extracting the shear due to gravitational lensing. Depending on the applications, the gravitational shear signal may be derived either by computing the power spectrum of the measured galaxy shapes, or by averaging the latter locally over small regions of the sky. We stress that MegaLUT, implemented as described below, aims at recovering ellipticities only, neglecting any shape parameter not used for shear studies. 

In the following, we will refer to \emph{observed} galaxy shapes when dealing with the shape of galaxies convolved by their PSF, as recorded on a detector. Note that these observed galaxies can be either real or simulated. In addition, we refer to \emph{sheared} galaxy shapes when dealing with the shape galaxies had prior to convolution by the atmospheric and instrumental PSF.

These \emph{sheared} galaxy shapes, and in particular the sheared ellipticities are what we are after. To recover them from the observed galaxy shapes, the proposed method needs some knowledge of the PSF either as a parametric model (e.g., Moffat, Gaussian or other more sophisticated profiles), or as a decomposition on a basis of vectors (e.g., Shapelets, Zernike or Hermite polynomials), or simply as an array of pixels, i.e., a sampled image of a star or a stack of stars. We assume that the PSF has already been estimated at best at the position of each galaxy in the survey, i.e., PSF interpolation is considered as a separate/solved problem. The way galaxies and PSFs are represented need not be the same, as long as the same representations are adopted for the real and the synthetic learning data.

Through this paper, we will use the notion of \emph{complex ellipticity}, common to shear studies, as defined in the GREAT10 challenge \citep{GREAT10Handbook2010}. This complex ellipticity, $e$, is linked to the \emph{elongation} $\epsilon$ and position angle $\theta$ of the objects, where  $\epsilon = a/b$, and $a$ and $b$ are respectively the semi-major and semi-minor axis of the light distribution isophotes:

\begin{eqnarray}
e & = & \frac{\epsilon - 1}{\epsilon + 1}\ e^{i 2 \theta} \\
   & = & |e| \left[ {\rm cos}(2 \theta) + i\ {\rm sin}(2 \theta) \right] \\
   & = & e_1 + i\ e_2
\end{eqnarray}

The factor 2 in the angular argument reflects shape invariance under rotation by 180$^\circ$. Complex ellipticity does \emph{not} encode the apparent size of an object.

Keeping the above in mind, MegaLUT consists of three steps: (1) generating a learning sample of simulated data, (2) building the lookup table (LUT) from this simulated data, (3) querying the LUT to recover the sheared galaxy shapes of the real data, i.e., the shapes of the (lensed) galaxies as they were before convolution by the PSF.

\begin{figure*}
\resizebox{\hsize}{!}{\includegraphics{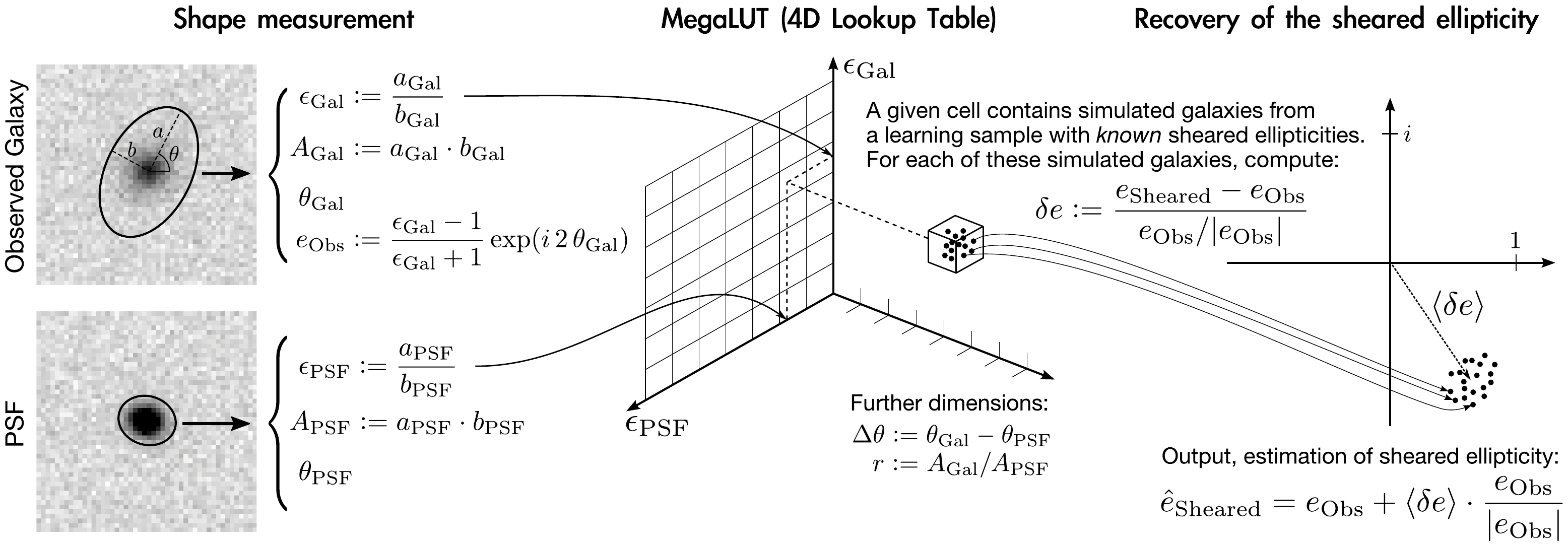}}
\caption{Structure of a MegaLUT query to recover the \emph{sheared} ellipticity, that is the ellipticity the galaxy had prior to the PSF convolution, from the observed shapes of the galaxy and the PSF. The shape measurement requires most of the computational time; it should be as precise as possible, but does not need to be \emph{accurate}, as MegaLUT cancels any biases in the coordinates.}
\label{fig_summary}
\end{figure*}

\subsection{Step 1 : generating the learning sample}

The first step is to build a learning sample of observed galaxies with \emph{known} sheared complex ellipticities $e_{\mathrm{Sheared}}$, and randomly associating a PSF to each of these galaxies. Properties like pixel size, noise characteristics, galaxy morphology and PSF profiles of this learning sample should be as close as possible to the data to be analysed. To build such a learning sample, where observed galaxies and PSFs are in the form of pixelized images, we adopt the following procedure:

\begin{enumerate}

\item Draw artificial sheared (i.e., weakly lensed) galaxies, and associated PSFs, on a fine pixel grid. The adopted pixel sampling for the artificial images should simply be chosen fine enough so that it does not influence the results, given the required precision. For each galaxy, store the sheared ellipticity $e_{\mathrm{Sheared}}$. For both the galaxies and the PSFs, randomly sample a broad range of radial profiles, apparent sizes, fluxes, ellipticities and orientations. This sampling can very well be uniform as long as it covers the full parameter space for real galaxies and PSFs.

\item Numerically convolve the galaxies with their associated PSFs.

\item Downsample the convolved galaxies and PSFs to match the pixel size of the real data.

\item Add realistic noise to the simulated images. The properties of the noise can easily be chosen to match that of the real data and may even include subtleties like cosmic rays and charge transfer inefficiency. In the present implementation the latter two effects are left out. 

\end{enumerate}

We now measure these simulated \emph{observed} galaxies with any given shape measurement algorithm, leading to a set of parameters, such as size, ellipticity, position angle and flux. The shape measurement algorithm can be seen as a black box; it should be precise, but not necessarily accurate, i.e., it should be as insensitive as possible to noise, while systematic biases in the measurements are acceptable. Those biases will be inherently cancelled by the method. 
We do the same with the simulated PSFs, leading to a set of associated PSF shape parameters. Note that the shape measurement algorithms applied to the galaxies and to the PSF need not be the same. 

At this point, the learning sample consists of an unordered collection of measured galaxy and PSF shape parameters, associated to the known underlying \emph{sheared} galaxy ellipticities.

\subsection{Step 2 : building the LUT}
\label{buildlut}

Next, we classify the galaxies from the learning sample according to these measured shape parameters. A given galaxy can be seen as a point in a multidimensional space, each dimension corresponding to one parameter, e.g., size, ellipticity, position angle and flux of the galaxy, as well as size, ellipticity and position angle of the PSF. Some of these observed parameters are clearly degenerate with respect to the sheared ellipticities. For example, the absolute sizes of the galaxy and the PSF are not required to recover the sheared ellipticity. What matters is the relative size of the galaxy with respect to the PSF. Following a similar theoretical argumentation, the measured fluxes of the sources seem a priori irrelevant. Note that in practice, the flux -- or signal-to-noise ratio $S/N$ -- of the galaxies and PSF stars could well be important, as it might bias the other measured shape parameters. For the specific application of MegaLUT described in this paper, we made use of a shape measurement whose biases do not significantly depend on the $S/N$ within the considered range, as we will show in Section \ref{shapemeasanalysis}. Therefore, we do indeed disregard the fluxes in the following.

Similarly, the PSF smearing should be invariant with respect to rotation on the plane of the sky\footnote{We neglect in this first approach possible rotation-dependent effects due to finite and square stamp size, pixel sampling, or charge transfer inefficiency.}. Hence, only the relative orientation between the PSF and the galaxy influences the correction for the PSF smearing. 

Both to accommodate for the parameter degeneracies and to minimize the dimensionality of the LUT, we reduce the parameter space to the following set of four less degenerate continuous \emph{coordinates}:
\begin{itemize}
\item $\epsilon_{\textrm{\,Gal}}$: the elongation of the galaxy
\item $\epsilon_{\textrm{\,PSF}}$: the elongation of the associated PSF
\item $r$: the size ratio between the galaxy and its associated PSF
\item $\Delta\theta$:  the relative orientation of the PSF with respect to the galaxy.
\end{itemize}
Each galaxy can now be represented as a point in this four-dimensional space. The classification consists of dividing this space into numerous hyperrectangles (``4-orthotopes'' in mathematical terms), that we call here \emph{cells}. This is easily done by individually splitting the full range of observed values of each of the above coordinates into a finite number of bins. The coordinates of any galaxy can then be univocally associated to a corresponding cell. 

Computationally, the four discretization functions that relate the continuous coordinate values to four indexes that identify a cell can be kept short and very fast. They consist mainly of a rounding operation. We have explored the situation in which all the cells have the same size, i.e. the binning of the coordinates is regular. While this is the simplest choice, it is by no means a required condition. 

Finally, to build the LUT, we distribute all galaxies of the learning sample among the corresponding cells. 
In each cell, the differences between the known \emph{sheared} and the measured \emph{observed} complex ellipticities $e_{\mathrm{Sheared}} - e_{\mathrm{Obs}}$ give an estimation for a simple additive correction to undo the smearing by the PSF. But the galaxies in a cell, and hence also these differences, have random position angles on the sky. To obtain complex ellipticity corrections that are rotation invariant we express their orientations with respect to the measured orientation of the galaxy. In mathematical terms, this corresponds to computing:
\begin{equation}
\delta e \mathrel{\mathop:}= \frac{e_{\mathrm{Sheared}} - e_{\mathrm{Obs}}}{e_{\mathrm{Obs}} / |e_{\mathrm{Obs}}|} \quad \mathrm{with\, }\, \delta e,\ e_{\mathrm{Sheared}},\ e_{\mathrm{Obs}} \in \mathbb{C}
\end{equation}
for each simulated galaxy.
These $\delta e$ can now be averaged within each cell. The LUT, as it will be used in the next step, thus consists of a multidimensional array of complex ellipticity corrections $\langle \delta e \rangle$. The standard deviation of the $\delta e$ within each cell can be used to express the uncertainty of these corrections.

\subsection{Step 3 : querying the LUT for real galaxies}

For each galaxy and PSF pair \emph{in the real data}, we measure the observed shape parameters using the exact same black box as applied on the learning sample. Hence any systematic errors inherent to the shape measurement are cancelled. The measured parameters are transformed into coordinates, and the coordinates are discretized into integer indexes of the LUT, again by the exact same simple functions that were used to build the LUT. Through a simple array indexing operation, the complex ellipticity correction $\langle \delta e \rangle$ can thus be directly read from the LUT. Finally, we obtain the estimation of the sheared ellipticity by applying the correction to the observed galaxy shape:
\begin{equation}
\hat{e}_{\mathrm{Sheared}} = e_{\mathrm{Obs}} + \langle \delta e \rangle \cdot
\frac{e_{\mathrm{Obs}}}{|e_{\mathrm{Obs}}|},
\end{equation}

where the multiplication by $e_{\mathrm{Obs}} / |e_{\mathrm{Obs}}|$ corresponds to a rotation of the correction by the measured orientation of the galaxy.

In the scope of cosmic shear surveys, the recovery of the sheared ellipticty has to be done for billions of galaxies, hence the simplicity and computational speed of this LUT query are crucial. Figure \ref{fig_summary} summarizes the procedure.

\section{Implementation for the GREAT10 challenge data}
\label{G10}

We applied MegaLUT to the GREAT10 Galaxy Challenge, descri\-bed in detail in the handbook and results papers \citep{GREAT10Handbook2010, G10results}. The challenge consists of recovering the shear power spectrum by measu\-ring the sheared ellipticities of 50 million simulated noisy galaxies placed on a rectangular grid. The shape of the PSF is variable across the field of view, but it is provided, at the position of each galaxy, both as a noisy pixelized stamp and under exact analytical form. For our implementation of MegaLUT, we have exclusively used the noisy PSF images.

In this section, we describe how we generated the learning sample for MegaLUT, and how we improved on the shape measurement since the end of the GREAT10 challenge. 

\subsection{Generation of the learning sample}

The GREAT10 Coordination Team has simulated the galaxy images to be analysed by superposing two exponential profiles of the form $\exp(-k R^{1/n})$, namely a disk ($n=4$) and a bulge ($n=1$), that may be misaligned with respect to each other. Then, before convolving them by the PSF, the shear signal was introduced by explicitly applying a distortion to the galaxy images \citep{GREAT10Handbook2010}.

To build our learning sample, we have chosen to \emph{directly} draw sheared galaxies using a single elliptical exponential profile with $n = 1.5$. Doing so we keep the generation of our learning sample as simple as possible, and show that a detailed knowledge of the GREAT10 simulation details is not required by our method. Furthermore, this simplification unambiguously links the true sheared ellipticities of our learning sample galaxies to their analytical form.

For the PSF, we use Moffat profiles with $\beta = 3$, i.e., the same profile that was used to generate the PSFs of GREAT10.

Before drawing the stamps for the learning set, we need to determine the ranges of sheared galaxy and PSF sizes and ellipticities so that the resulting measured shape parameters, and thus the 4 coordinates of MegaLUT as described in section \ref{buildlut}, cover the values required to process the GREAT10 data. In practice, we therefore first run a shape measurement algorithm on the GREAT10 data, and then empirically adjust the input parameter ranges of the learning simulations so that the observed characteristics ($a_\mathrm{Gal}$, $a_\mathrm{PSF}$, $\epsilon_\mathrm{Gal}$, $\epsilon_\mathrm{PSF}$) match those of the GREAT10 stamps.
Position angles of the galaxies and PSFs follow a uniform distribution across all possible orientations. The signal-to-noise ratio ($S/N$) is simply kept constant, for both the galaxies and the PSFs, to the fiducial $S/N$ of the GREAT10 data. The centroid positions of the galaxy and PSF profiles within the stamps are randomized by a uniformly distributed scatter of $\pm1$ large pixel in each direction.

We now draw, and then convolve, these galaxy and PSF profiles on fine pixels, $4$ times smaller than the GREAT10 pixels (i.e., 16 times in area). We bin the pixels $4 \times 4$ to match the GREAT10 sampling and we add a simple Gaussian noise with $\sigma = 1$ ADU (in the same flux scale as GREAT10) to  the convolved galaxy and to the PSF images. This is well representative of the ``sky-limited'' acquisition regime. Taken individually, the resulting stamps are indistinguishable from the GREAT10 data in terms of the proposed shape measu\-rement and visual inspection.

\subsection{Shape measurement methods}
\label{g10shapemeas}

In the scope of the GREAT10 challenge, we have compared two fast shape measurement methods, both based on the computation of the 2nd order moments of the light distribution. Here we briefly describe them and assess their precision.

\subsubsection{Masked 2nd moments + denoising (hereafter MMD)}
\label{mmd}

To optimally include the shape measurement in our workflow and test the feasibility of the proposed method, we have, in a first step, implemented our own shape measurement. It sequentially processes the pairs of galaxies and PSFs, stamp by stamp, and can be summarized as follows :

\begin{enumerate}

\item Denoise the stamp, using hard thresholding of the first and second level of its Haar wavelet coefficients. 

\item Build a boolean isophotal mask for the denoised stamp, selecting only those pixels whose values are above a certain fraction of the maximum value.

\item Compute the barycenter, and the centered 2nd order moments of the resulting (i.e. masked and denoised) stamps.

\item Transform the 2nd order moments into an orientation $\theta$ and a semi-major and semi-minor axes $a$ and $b$. In doing this, we use the same formalism as in the SExtractor package  \citep{2010ascl.soft10064B}.
\end{enumerate}

The denoising step is important to increase the precision of the measured parameters and also because it smoothes the contour of the isophotal mask. As MegaLUT cancels biases of the shape measurement step, we can make use of a rather strong denoising, even if the latter does not completely preserve object shapes.

We have submitted MegaLUT only in combination with this first MMD shape measurement method to the GREAT10 challenge leaderbord.

\subsubsection{SExtractor windowed 2nd moments (SEWIN)}
\label{sewin}

The widely used SExtractor software \citep{sextractor} implements ``windowed'' measurements of the centroids and 2nd order moments. The computation of the latter are similar to the basic 2nd order moments, except that the pixel values are weighted, in a similar way to which it is done in KSB, by an adaptive circular Gaussian window\footnote{The windowed parameters are described in the SExtractor manual by E. Bertin, available at \url{http://www.astromatic.net/software/sextractor}}. While these windowed parameters can be significantly biased with respect to the basic ones, they are far less sensitive to noise in the input images.

As an alternative to the simple MMD shape measurement, we thus include a second shape measurement based solely on the latest SExtractor (version 2.8.6) in our analysis. To retain its advantage in computational speed, simplicity, and reproducibility, we do not combine it with prior denoising of the images.

We have considered this second shape measurement method only after the GREAT10 challenge deadline. Thus its results are not included in \citet{G10results}.

\subsubsection{Analysis of the shape measurement precision}
\label{shapemeasanalysis}

For MegaLUT to deliver precise results, the scatter of ellipticity corrections within each cell should be as small as possible (e.g., see Fig.~\ref{fig_summary}). This scatter has three sources:

\begin{enumerate}

\item The precision of the shape measurement, i.e., the sensitivity of the coordinates to noise in the images. An imprecise shape measurement will randomly allocate the learning sample galaxies to the wrong cells. Additionally, these random errors will also influence the querying the LUT. Note that in practice, galaxies and PSF images are sampled on a discrete pixel grid, resulting in an inherent limit in precision for any shape measurement.

\item The reduction of the multidimensional parameter space to a limited number of coordinates. A given choice of coordinates effectively marginalizes over all parameter dependencies not explicitly included in the chosen coordinates. If for instance the measurement of an elongation depends on the signal to noise ratio ($S/N$) of a source, and the LUT does not discriminate according to $S/N$, stamps differing only by their $S/N$ may be allocated to different cells.

\item The actual variation of the ellipticity correction within the finite size of the cells, due to the continuous evolution of the ellipticity corrections, $\delta e$, in the parameter space. This effect is inherent to the method, but can be easily adressed by choosing a sufficiently fine sampling of the parameter space. For the sampling used in our implementation, this source of scatter is insignificant compared to the first two points.

\end{enumerate}

We evaluate the precision of the two shape measurement methods presented in the previous sections by running them on 400 realizations of a single si\-mu\-lated galaxy and an associated PSF. The corresponding stamps are drawn using the same light profiles as for the learning sample, but we keep all parameters of the profiles constant, by setting them to typical values representative of the GREAT10 data. Only the noise realization and the scatter in centroid positions differ between these simulated stamps.

\begin{figure}
\resizebox{\hsize}{!}{\includegraphics{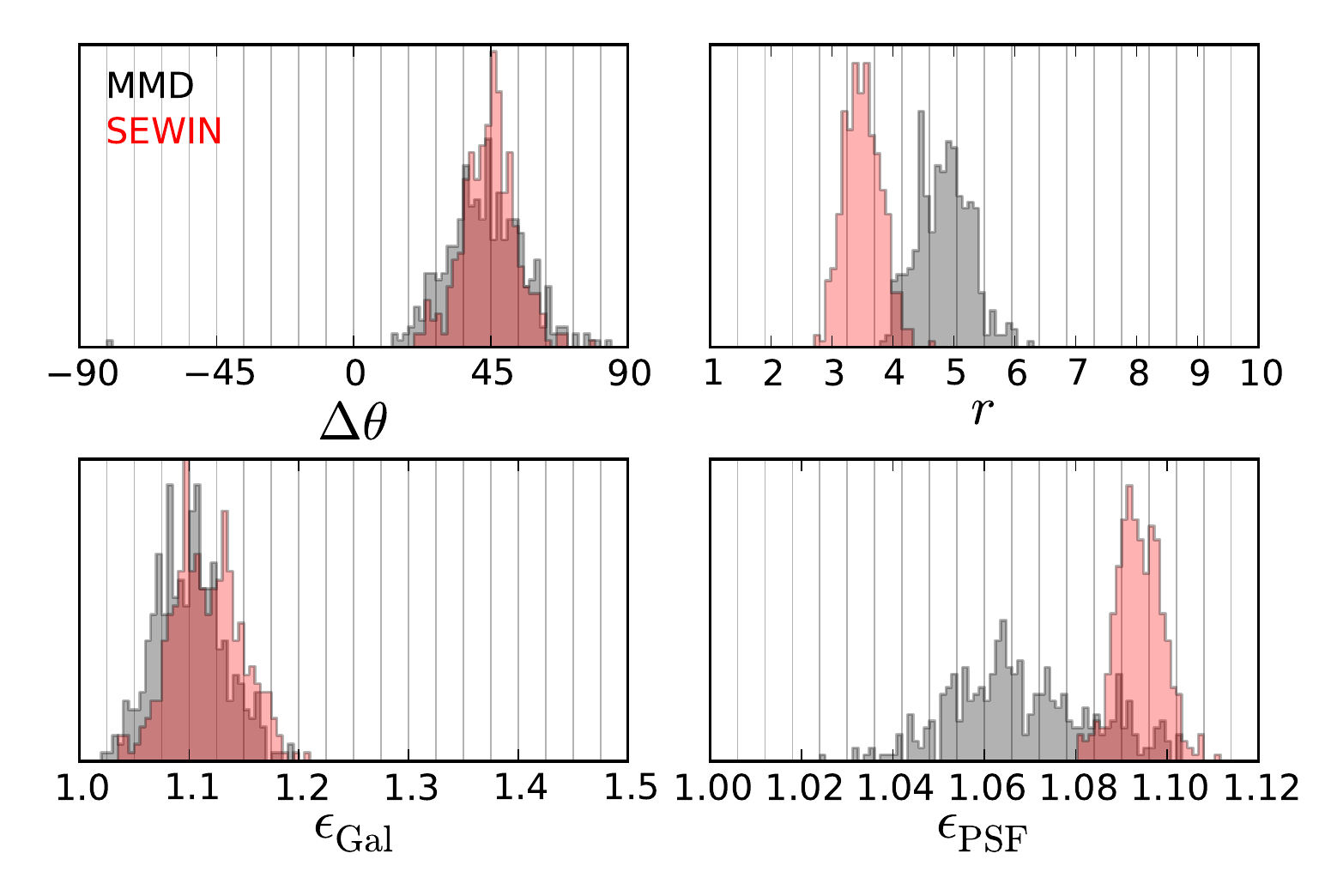}}
\caption{Comparison of the precision of the shape measurements methods used in this work, obtained by running them on 400 noisy realizations of always the same galaxy and PSF pair. MMD results are shown in grey, SEWIN in red. The shape parameters measured by MMD are rescaled so that their variance can be equitably compared to the variance of the SEWIN ones (see text). The vertical lines indicate the 20 bins in each coordinate, as used for all applications to GREAT10 described in this paper. See Fig. \ref{fig_summary} for a description of the coordinates.}
\label{fig_MMDvsSEWIN}
\end{figure}

The histograms for the four coordinates obtained through the two methods are shown in Fig. \ref{fig_MMDvsSEWIN}.  In this plot, each coordinate $c$ obtained from MMD has been linearly rescaled ($c' = m \cdot c$) so that its variance can be equitably compared to the variance of the SEWIN coordinate. Indeed, the different ways of masking and weighting the second order moments yield significantly different raw shape parameters; for instance, the elongations measured by SEWIN are systematically about half of the elongations from MMD. For each coordinate, the scaling factor $m$ is chosen so that the range of coordinates computed for the full learning sample by the two techniques robustly overlaps. Note that this rescaling is only required for the comparative study of Fig. \ref{fig_MMDvsSEWIN}.

Discrepancies in accuracy (i.e. positions of the peaks) of the techniques is not a concern, as MegaLUT corrects for bias using the learning sample; the peaks should simply be as narrow as possible. The SEWIN method clearly evinces a higher precision than the simple MMD that was submitted to the GREAT10 challenge. This is especially true for the measurement of the elongation of the PSF.

The width of the histograms in Fig. \ref{fig_MMDvsSEWIN} gives the \emph{resolution} of the shape measurement for data very similar to GREAT10. The bin size used to discretize the coordinates of the LUT cells can now be chosen fine enough to avoid any significant degradation of this resolution.

\begin{figure}
\resizebox{\hsize}{!}{\includegraphics{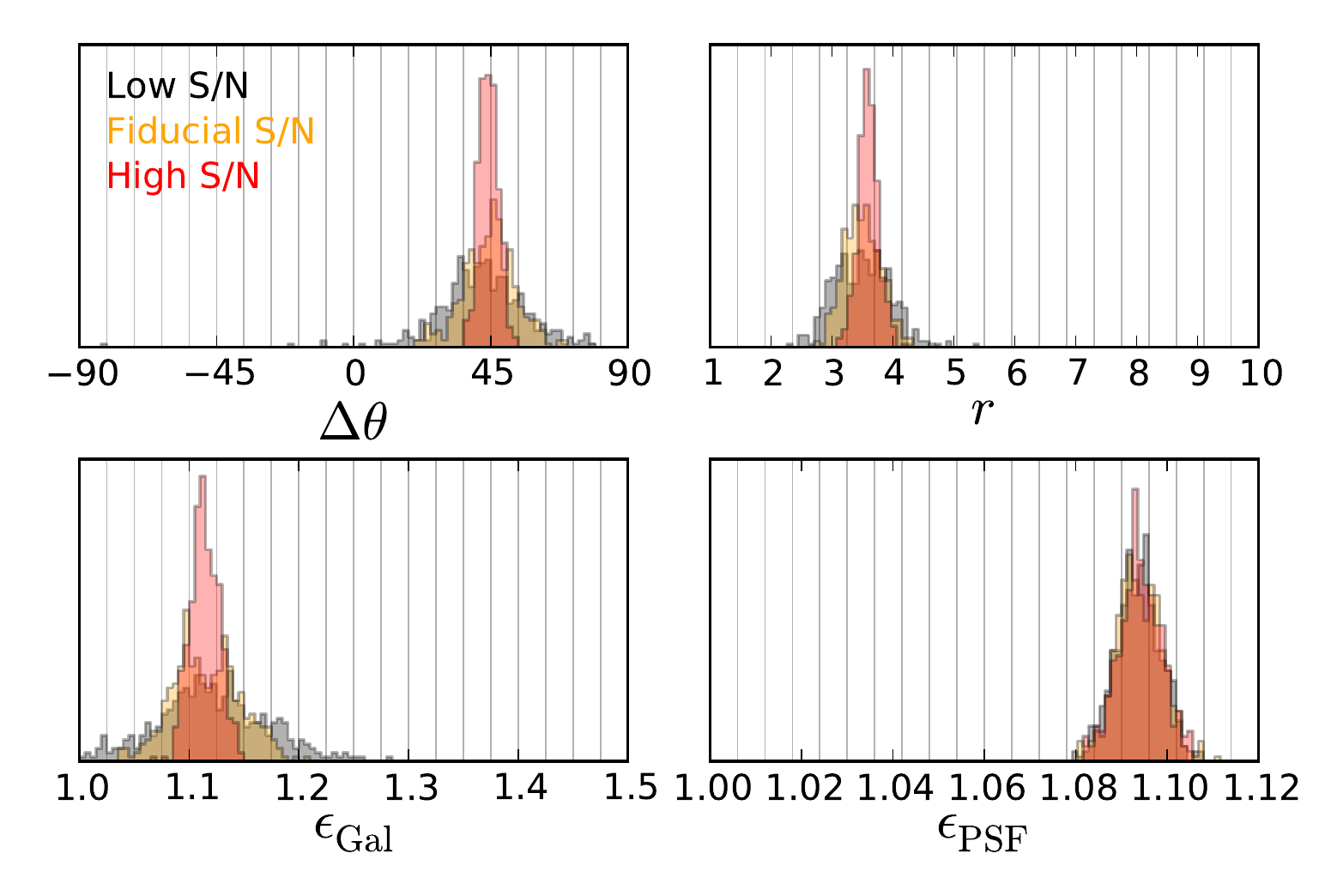}}
\caption{Sensitivity of SEWIN coordinates to the 3 different signal-to-noise ratios of GREAT10, for a typical galaxy and PSF. There is no observable bias of the shape measurements with changing $S/N$. Therefore we do not include the $S/N$ as a dimension of the LUT.}
\label{fig_SN}
\end{figure}

To evaluate the importance of the second source of scatter, that is the marginalization over potentially discriminating parameters, we process in a similar way. In Fig.~\ref{fig_SN}, we compare the SEWIN measurements for the 3 different signal-to-noise ratios encountered in the GREAT10 data. We observe that the centroids of all 4 coordinates, as obtained from SExtractor, are not significantly affected by the $S/N$, at least within the range of $S/N$ explored in GREAT10. This is a remarkable property of SExtractor's windowed moments, hence it is justified not to include the $S/N$ as a coordinate in the LUT. Naturally, we do observe an increase in the \emph{variance} of the coordinates with decreasing $S/N$; such a lack of precision inevitably degrades the shear signal, whatever be the accuracy of the correction.

\begin{figure}
\resizebox{\hsize}{!}{\includegraphics{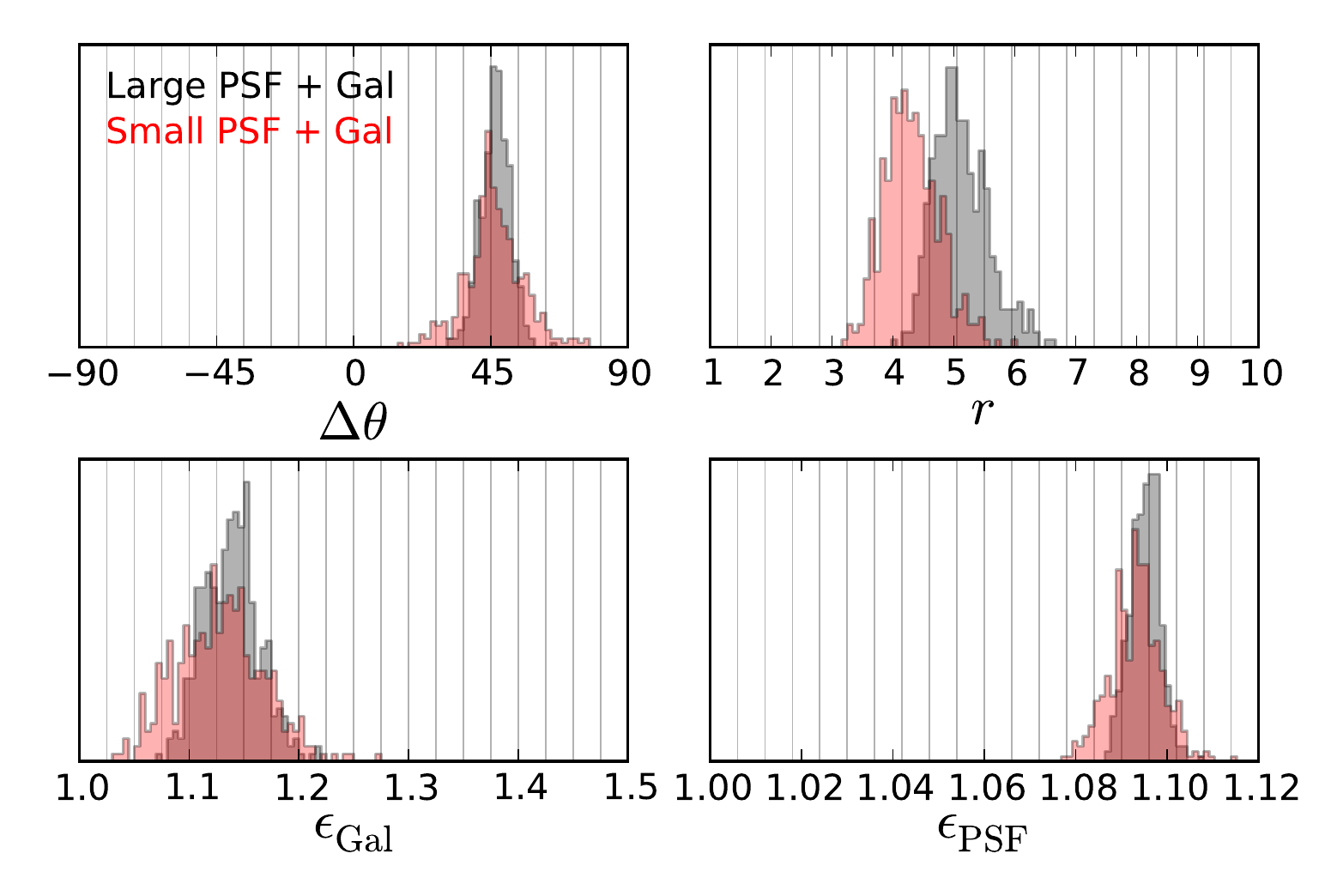}}
\caption{Sensitivity of SEWIN coordinates to the size of the PSF and galaxy pair. The distinction between the small (red) and large (blue) couples is a factor 1.8 in full width at half maximum of both the PSF and the galaxy, covering the range from the smallest to the widest PSFs in GREAT10.
For the present MegaLUT implementation, the measurement of the size ratio $r$ should ideally \emph{not} depend on this rescaling.}
\label{fig_size}
\end{figure}

Figure~\ref{fig_size} illustrates a similar analysis, but varying the size of the galaxy and PSF, instead of the $S/N$. The coordinate $r$ discriminates stamps by their galaxy-to-PSF area \emph{ratio}. For an analytical convolution, this ratio would not be affected by rescaling both the galaxy and the PSF by the same factor. However, as a consequence of the pixelization, the histogram for the values of $r$ slightly depends on the galaxy and PSF scale. At the cost of adding one more dimension to the LUT, this limitation may be adressed by including both the galaxy and the PSF size as coordinates to the LUT, instead of their ratio.

Aside from the mentioned sources of scatter within the cells, the ellipticity corrections can also be \emph{biased}, if the learning sample is not representative enough of the galaxies and PSFs to be analyzed. Our method is indeed a machine learning method. As such, the quality of the training set is important. We discuss this source of error in Section \ref{discussion}.

\subsection{Building the LUT}

Given the resolution of the two considered shape measurement methods (see figure \ref{fig_MMDvsSEWIN}), we have chosen, for all our applications to GREAT10, to use a regular sampling of 20 bins in each of the 4 coordinates, yielding $20^4 = 160'000$ cells. As expected, further increasing this sampling did not improve the performance of the algorithm.

\begin{figure}
\resizebox{\hsize}{!}{\includegraphics{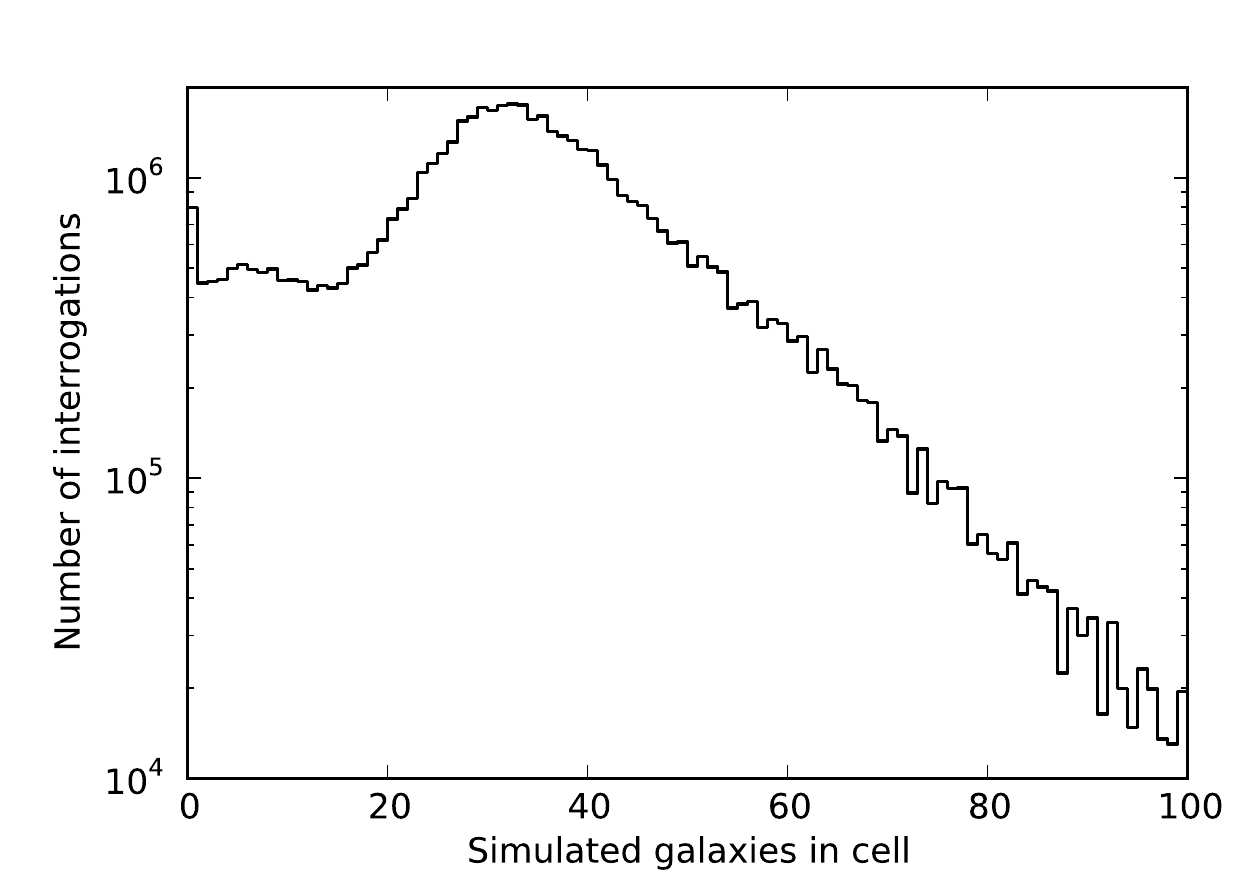}}
\caption{Histogram of the number of learning sample galaxies in the cells as encountered by the 50 million queries to the LUT. The cells of the LUT that are most queried by the GREAT10 data are sufficiently filled by the learning sample. Using a learning sample of 9 million galaxies, we find that only 5\% of the queries found less than 5 learning galaxies in their cells. Note that for those queries falling on an empty cell, our implementation returns the mean correction from the cells neighbors.}
\label{fig_queryhist}
\end{figure}

For our submission of MegaLUT using the MMD shape measurement to the GREAT10 challenge, we built a learning sample of 2.1 million galaxies. Since then, we increased this number to 9 millions, without changing neither the profiles nor the parameter distributions. As illustrated in figure \ref{fig_queryhist}, this number is large enough to sufficiently fill the required cells of the LUT. But in fact, we observe that the GREAT10 scores achieved do not significantly decrease when using only our initial learning sample of 2.1 million galaxies.

\section{Results on the GREAT10 challenge data}
\label{G10results}

We participated in the GREAT10 challenge by combining MegaLUT with the MMD shape measurement (Section \ref{mmd}), reaching an encouraging quality factor $Q$ of 69.2 \citep{G10results}.

With the SExtractor-based SEWIN shape measurement (Section \ref{sewin}) -- implemented after the challenge deadline, and thus not in the official leaderboard -- MegaLUT reaches a $Q$ factor of 104, without power spectrum denoising or training. This score is competitive with the results achieved by the best ellipticity catalog submission to GREAT10.

\begin{figure*}
\resizebox{\hsize}{!}{\includegraphics{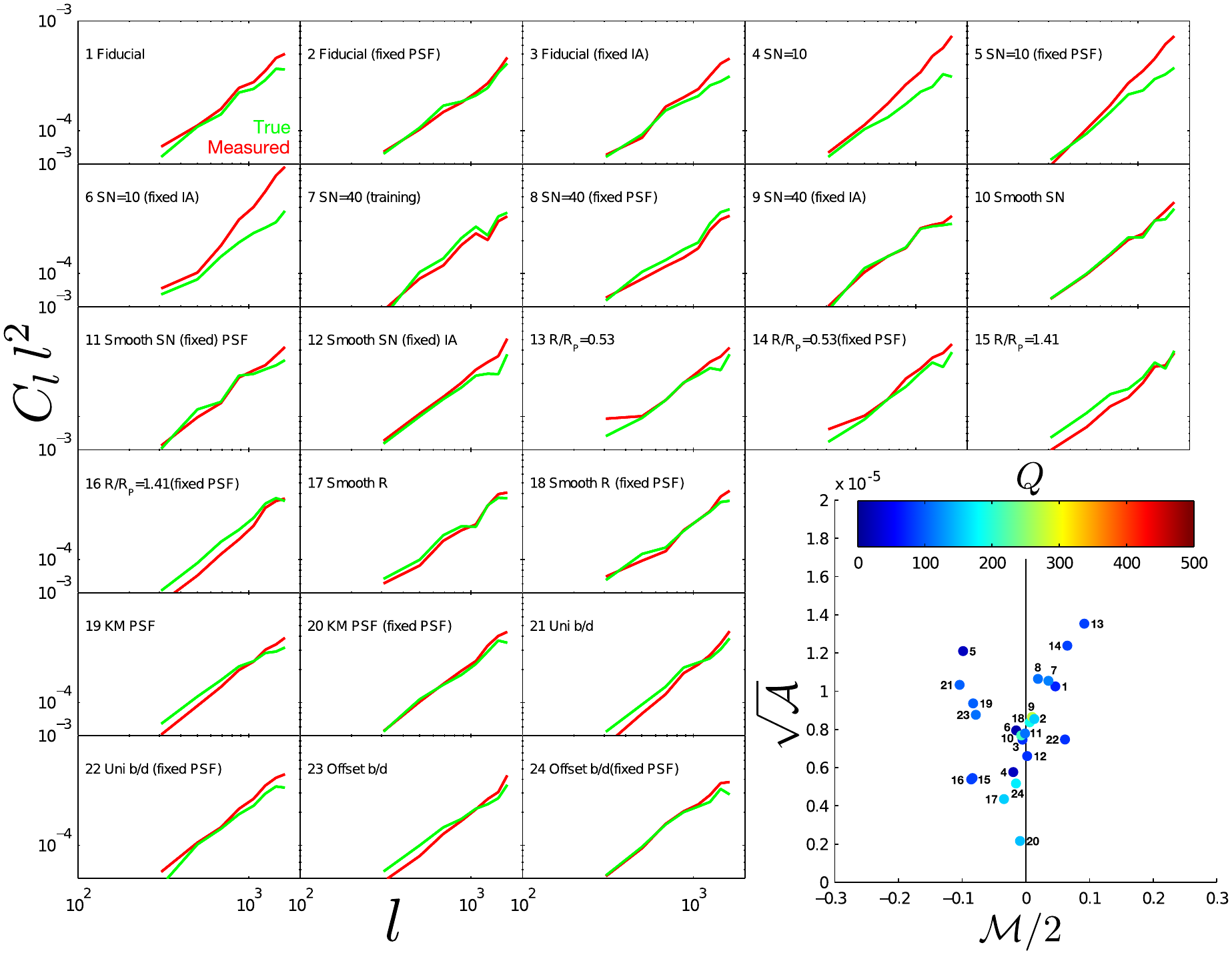}}
\caption{
Power spectra of the sheared ellipticities as obtained by MegaLUT + SEWIN for each set of the challenge. This figure can be directly compared with those from all GREAT10 submissions described in \citet{G10results}. The red lines represent the \emph{measured} shear power, obtained without the denoising term, while the green lines represent the \emph{true} shear power. The inset gives the metrics $\mathcal{M}/2$, $\sqrt{\mathcal{A}}$ and $Q$ (without denoising or training) for each set.
}
\label{fig_PSG10}
\end{figure*}

The achieved values of all GREAT10 metrics obtained using the two shape measurements are presented in Table \ref{Qtable}. We observe that the SEWIN shape measurement substantially improves (i.e., reduces) both the one-point ($m$, $c$) and power spectrum ($\mathcal{M}/2$, $\sqrt{\mathcal{A}}$) bias estimates. Performance details of MegaLUT + SEWIN, for each set of the GREAT10 data, are displayed in Figure \ref{fig_PSG10}.

\setlength{\tabcolsep}{3pt}
\begin{table}[htdp]
\caption{The quality factors $Q$ and further GREAT10 metrics obtained by MegaLUT in combination with the two discussed shape measurement methods MMD and SEWIN.}
\renewcommand{\arraystretch}{1.35}
\begin{center}
\begin{tabular}{lccccccc}
\hline
\hline
Method & $Q$ & $m$ & $c/10^{-4}$ & $\mathcal{M}/2$ & $\sqrt{\mathcal{A}}/10^{-4}$ \\
\hline
MegaLUT+MMD    & $\069.17$ & $-0.27$ & $-0.550$ & $-0.1831$ & $0.1311$ \\
MegaLUT+SEWIN & $104.14$ & $ -0.15$ & $-0.057$ & $\;\;\,0.0119$ & $0.0819$\\
\hline
\end{tabular}
\tablefoot{All values are computed using the same analysis code as for the GREAT10 results paper. The description of these metrics can be found in \citet{G10results}. The entry MegaLUT + MMD refers to the submission ``MegaLUTsim2.1 b20'' in \citet{G10results}.
}
\end{center}
\label{Qtable}
\end{table}

We implemented MegaLUT in pure PYTHON in a few hundred lines of code\footnote{The code is available at \url{http://lastro.epfl.ch/megalut}}. Using the SExtractor shape measurement, the whole process of detecting, characterizing the galaxy/PSF pairs, and querying the LUT takes less than 3 milliseconds per galaxy on an AMD Opteron 2216 2.4 GHz CPU. Given the competitive quality metrics, this makes MegaLUT a very efficient solution to the PSF correction problem.

\section{Discussion}
\label{discussion}

MegaLUT splits the measurement process of sheared ellipticities into two distinct parts : the shape measurement itself, and the subsequent ellipticity correction by a simple form of supervised learning. Limitations of the shape measurement algorithms in the presence of noise as well as the finite sampling and dimensionality of the LUT are practical error sources. They are discussed in Section \ref{shapemeasanalysis}. We recall that even if our empirical method corrects for biases on the estimated sheared ellipticites, the remaining variance in this output will degrade the weak lensing signal, and in particular bias the shear power spectrum.

The remaining and more fundamental sources of error concern the discrepancies between the learning sample and the data to be analyzed. In this paper, we have kept the learning sample as elementary as possible, using a single simply parametrized profile for the galaxies. All the free parameters describing the generation of these learning galaxies and the associated PSFs, such as size and ellipticity, directly relate to coordinates of the LUT. Assuming a perfect shape measurement and noiseless data, two galaxy-PSF pairs from the learning sample would get attributed the same coordinates only if the pairs are virtually identical, except for their absolute orientation and size. As a consequence, for such a simple parametrization of the learning sample, the actual distributions of parameters used to generate the learning sample do not act as priors of the method. Indeed, the queried ellipticity corrections don't depend on these distributions, as long as the LUT gets sufficiently filled with learning data in all required cells.

But clearly, real galaxies do not follow smooth exponential light profiles. Instead, their possibly multiple components follow a variety of slopes, contain asymmetries, and may well not come isolated. Any employed shape measurement is sensitive to these substructures. Therefore, a machine learning approach like MegaLUT will yield biased results if it is not trained on realistic galaxies. How can we deal with this necessity for a realistic learning sample ?

Let us note that for a real galaxy there is no longer a natural and unambiguous definition of ellipticity as we have for the simple smooth profiles with perfectly elliptical isophotes. The ellipticity of a real galaxy must be defined through a measurement on the image. Hence, to combine MegaLUT with a more detailed and realistic learning sample, one can easily \emph{measure} the sheared ellipticities of the simulated galaxies before the convolution by the PSF and the addition of noise.  This procedure allows to use simulated learning galaxies with arbitrary substructure, and also to shear them in a well controlled way once they have been drawn on a pixel grid.
When such a detailed learning sample is built, the distribution of parameters describing the generation of galaxy substructure (e.g. light profiles, clumps, companions) would influence the distribution of ellipticity corrections inside the LUT cells. They would thus effectively act as priors on the method, to be chosen according to the population of galaxies to be analyzed. Ideally such simulated data should be 
cosmology-independent and blind, to avoid confirmation bias effects.

Furthermore, such an increase in details of the learning sample represents an opportunity for more sophisticated shape measurement methods to test the benefits of additional characterizations of the galaxies and PSF, as for example an estimation of the radial slope of the light distributions. If consequently the desired number of coordinates or cells of the LUT increases significantly, the memory requirements and CPU time for the generation of enough learning data might become a limitation. In any case, the brute-force LUT with manually chosen coordinates could be replaced by a fast interpolation across sparse data in a large parameter space, for instance by using an artificial neural network like those employed by \cite{gruen2010}.

\section{Conclusion}
\label{conclusion}

In this paper we have presented MegaLUT, a new method to correct galaxy shape measurements from smearing by the instrumental and atmospheric PSF. We list below a summary of the advantages of our method.

\begin{enumerate}

\item MegaLUT is empirical. It does not need to rely on a specific shape measurement method or shape definition, and does not require the shape measurement to be accurate (bias is tolerated) as long as it is precise (low variance). The shape measurement itself can be considered as an interchangeable black box.

\item As a consequence, MegaLUT can well be combined with existing shape measurements techniques, in particular it can make use of strong image denoising to increase the shape precision, even if the denoising itself introduces biases in the measured parameters.

\item Each galaxy is processed individually, hence MegaLUT is independent from the spatial power spectrum of the shear field or the PSF variations.

\item The total computational cost of the analysis of a galaxy and its corresponding PSF is dominated by the shape measurement process, as the shape \emph{correction} essentially reduces to a simple array indexing operation. When combined with an efficient shape measurement, MegaLUT is fast, with a total processing time of a few milliseconds per galaxy, on an ordinary CPU. 

\end{enumerate}

By applying this method to the GREAT10 challenge \citep{GREAT10Handbook2010, G10results}, we have shown that its results are well competitive ($Q$ = 104) with the best submitted methods, despite a truly simplistic learning sample and the lack of additional corrections for bias at the level of the shear power spectra. Like for any machine learning technique, once the technical aspects are well controlled, it's ultimately the quality of this learning sample that limits the performance of the shape measurement itself. To obtain the best possible shape estimates for real weak lensing observations, a more representative learning sample might be required. We have discussed in Section \ref{discussion} how a learning sample containing arbitrarily realistic galaxies and PSFs could easily be used. In particular, such a learning sample can be build directly using high-resolution observations, like Hubble Space Telescope images.

\begin{acknowledgements}
This work is supported by the Swiss National Science Foundation (SNSF).
We thank the GREAT10 Coordination Team for organizing the stimulating challenge and sharing the quality factor calculation codes. GREAT10 was sponsored by a EU FP7 PASCAL 2 challenge grant. TDK was supported by a Royal Society University Research Fellowship. We would also like to thank the anonymous referee for her/his beneficial comments.
\end{acknowledgements}

\bibliographystyle{aa}
\bibliography{biblio}

\begin{thebibliography}{27}
\expandafter\ifx\csname natexlab\endcsname\relax\def\natexlab#1{#1}\fi

\bibitem[{{Bacon} {et~al.}(2000){Bacon}, {Refregier}, \& {Ellis}}]{Bacon2000}
{Bacon}, D.~J., {Refregier}, A.~R., \& {Ellis}, R.~S. 2000, \mnras, 318, 625

\bibitem[{{Bertin} \& {Arnouts}(1996{\natexlab{a}})}]{sextractor}
{Bertin}, E. \& {Arnouts}, S. 1996{\natexlab{a}}, \aaps, 117, 393

\bibitem[{{Bertin} \& {Arnouts}(1996{\natexlab{b}})}]{2010ascl.soft10064B}
{Bertin}, E. \& {Arnouts}, S. 1996{\natexlab{b}}, in Astrophysics Source Code
  Library, record ascl:1010.064, 10064

\bibitem[{{Bolton} {et~al.}(2008){Bolton}, {Burles}, {Koopmans}, {Treu},
  {Gavazzi}, {Moustakas}, {Wayth}, \& {Schlegel}}]{Bolton2008}
{Bolton}, A.~S., {Burles}, S., {Koopmans}, L.~V.~E., {et~al.} 2008, \apj, 682,
  964

\bibitem[{{Coe} {et~al.}(2010){Coe}, {Ben{\'{\i}}tez}, {Broadhurst}, \&
  {Moustakas}}]{Coe2010}
{Coe}, D., {Ben{\'{\i}}tez}, N., {Broadhurst}, T., \& {Moustakas}, L.~A. 2010,
  \apj, 723, 1678

\bibitem[{{Courbin} {et~al.}(2012){Courbin}, {Faure}, {Djorgovski},
  {R{\'e}rat}, {Tewes}, {Meylan}, {Stern}, {Mahabal}, {Boroson}, {Dheeraj}, \&
  {Sluse}}]{Courbin2012}
{Courbin}, F., {Faure}, C., {Djorgovski}, S.~G., {et~al.} 2012, \aap, 540, A36

\bibitem[{{Faure} {et~al.}(2008){Faure}, {Kneib}, {Covone}, {Tasca},
  {Leauthaud}, {Capak}, {Jahnke}, {Smolcic}, {de la Torre}, {Ellis},
  {Finoguenov}, {Koekemoer}, {Le Fevre}, {Massey}, {Mellier}, {Refregier},
  {Rhodes}, {Scoville}, {Schinnerer}, {Taylor}, {Van Waerbeke}, \&
  {Walcher}}]{Faure2008}
{Faure}, C., {Kneib}, J.-P., {Covone}, G., {et~al.} 2008, \apjs, 176, 19

\bibitem[{{Gruen} {et~al.}(2010){Gruen}, {Seitz}, {Koppenhoefer}, \&
  {Riffeser}}]{gruen2010}
{Gruen}, D., {Seitz}, S., {Koppenhoefer}, J., \& {Riffeser}, A. 2010, \apj,
  720, 639

\bibitem[{{Hu}(1999)}]{Hu1999}
{Hu}, W. 1999, \apjl, 522, L21

\bibitem[{{Kacprzak} {et~al.}(2012){Kacprzak}, {Zuntz}, {Rowe}, {Bridle},
  {Refregier}, {Amara}, {Voigt}, \& {Hirsch}}]{Kacprzak2012}
{Kacprzak}, T., {Zuntz}, J., {Rowe}, B., {et~al.} 2012, arXiv:1203.5049

\bibitem[{{Kaiser} {et~al.}(1995){Kaiser}, {Squires}, \&
  {Broadhurst}}]{KSB1995}
{Kaiser}, N., {Squires}, G., \& {Broadhurst}, T. 1995, \apj, 449, 460

\bibitem[{{Kaiser} {et~al.}(2000){Kaiser}, {Wilson}, \& {Luppino}}]{Kaiser2000}
{Kaiser}, N., {Wilson}, G., \& {Luppino}, G.~A. 2000, arXiv:astro-ph/0003338

\bibitem[{{Kitching} {et~al.}(2011){Kitching}, {Amara}, {Gill}, {Harmeling},
  {Heymans}, {Massey}, {Rowe}, {Schrabback}, {Voigt}, {Balan}, {Bernstein},
  {Bethge}, {Bridle}, {Courbin}, {Gentile}, {Heavens}, {Hirsch}, {Hosseini},
  {Kiessling}, {Kirk}, {Kuijken}, {Mandelbaum}, {Moghaddam}, {Nurbaeva},
  {Paulin-Henriksson}, {Rassat}, {Rhodes}, {Sch{\"o}lkopf}, {Shawe-Taylor},
  {Shmakova}, {Taylor}, {Velander}, {van Waerbeke}, {Witherick}, \&
  {Wittman}}]{GREAT10Handbook2010}
{Kitching}, T., {Amara}, A., {Gill}, M., {et~al.} 2011, Ann.Appl.Stat., 5, 2231

\bibitem[{{Kitching} {et~al.}(2012){Kitching}, {Balan}, {Bridle}, {Cantale},
  {Courbin}, {Gentile}, {Gill}, {Harmeling}, {Heymans}, {Hirsch}, {Kacprzak},
  {Kirkby}, {Margala}, {Massey}, {Melchior}, {Nurbaeva}, {Patton}, {Rhodes},
  {Rowe}, {Taylor}, {Tewes}, {Viola}, {Witherick}, {Voigt}, {Young}, \&
  {Zuntz}}]{G10results}
{Kitching}, T.~D., {Balan}, S.~T., {Bridle}, S., {et~al.} 2012, arXiv:1202.5254

\bibitem[{{Kitching} {et~al.}(2008){Kitching}, {Miller}, {Heymans}, {van
  Waerbeke}, \& {Heavens}}]{Kitching2008}
{Kitching}, T.~D., {Miller}, L., {Heymans}, C.~E., {van Waerbeke}, L., \&
  {Heavens}, A.~F. 2008, \mnras, 390, 149

\bibitem[{{Kuijken}(2006)}]{Kuijken2006}
{Kuijken}, K. 2006, \aap, 456, 827

\bibitem[{{Laureijs} {et~al.}(2011){Laureijs}, {Amiaux}, {Arduini},
  {Augu{\`e}res}, {Brinchmann}, {Cole}, {Cropper}, {Dabin}, {Duvet}, {Ealet},
  \& et~al.}]{Laureijs2011}
{Laureijs}, R., {Amiaux}, J., {Arduini}, S., {et~al.} 2011, arXiv:1110.3193

\bibitem[{{Limousin} {et~al.}(2007){Limousin}, {Richard}, {Jullo}, {Kneib},
  {Fort}, {Soucail}, {El{\'{\i}}asd{\'o}ttir}, {Natarajan}, {Ellis}, {Smail},
  {Czoske}, {Smith}, {Hudelot}, {Bardeau}, {Ebeling}, {Egami}, \&
  {Knudsen}}]{Limousin2007}
{Limousin}, M., {Richard}, J., {Jullo}, E., {et~al.} 2007, \apj, 668, 643

\bibitem[{{Maoli} {et~al.}(2001){Maoli}, {Van Waerbeke}, {Mellier},
  {Schneider}, {Jain}, {Bernardeau}, {Erben}, \& {Fort}}]{Maoli2001}
{Maoli}, R., {Van Waerbeke}, L., {Mellier}, Y., {et~al.} 2001, \aap, 368, 766

\bibitem[{{Melchior} \& {Viola}(2012)}]{MelchiorViola2012}
{Melchior}, P. \& {Viola}, M. 2012, arXiv:1204.5147

\bibitem[{{Miller} {et~al.}(2007){Miller}, {Kitching}, {Heymans}, {Heavens}, \&
  {van Waerbeke}}]{Miller2007}
{Miller}, L., {Kitching}, T.~D., {Heymans}, C., {Heavens}, A.~F., \& {van
  Waerbeke}, L. 2007, \mnras, 382, 315

\bibitem[{{Refregier}(2003)}]{Refregier2003a}
{Refregier}, A. 2003, \mnras, 338, 35

\bibitem[{{Refregier} \& {Bacon}(2003)}]{Refregier2003b}
{Refregier}, A. \& {Bacon}, D. 2003, \mnras, 338, 48

\bibitem[{{Refregier} {et~al.}(2012){Refregier}, {Kacprzak}, {Amara}, {Bridle},
  \& {Rowe}}]{Refregier2012}
{Refregier}, A., {Kacprzak}, T., {Amara}, A., {Bridle}, S., \& {Rowe}, B. 2012,
  arXiv:1203.5050

\bibitem[{{Shan} {et~al.}(2012){Shan}, {Kneib}, {Tao}, {Fan}, {Jauzac},
  {Limousin}, {Massey}, {Rhodes}, {Thanjavur}, \& {McCracken}}]{Shan2012}
{Shan}, H., {Kneib}, J.-P., {Tao}, C., {et~al.} 2012, \apj, 748, 56

\bibitem[{{Van Waerbeke} {et~al.}(2000){Van Waerbeke}, {Mellier}, {Erben},
  {Cuillandre}, {Bernardeau}, {Maoli}, {Bertin}, {McCracken}, {Le F{\`e}vre},
  {Fort}, {Dantel-Fort}, {Jain}, \& {Schneider}}]{vanWaerbeke2000}
{Van Waerbeke}, L., {Mellier}, Y., {Erben}, T., {et~al.} 2000, \aap, 358, 30

\bibitem[{{Wittman} {et~al.}(2000){Wittman}, {Tyson}, {Kirkman},
  {Dell'Antonio}, \& {Bernstein}}]{Wittman2000}
{Wittman}, D.~M., {Tyson}, J.~A., {Kirkman}, D., {Dell'Antonio}, I., \&
  {Bernstein}, G. 2000, \nat, 405, 143

\end{thebibliography}
\end{document}